# Modeling Cultural Dynamics

**Liane Gabora**

University of British Columbia
Okanagan campus, 3333 University Way
Kelowna BC, V1V 1V7, CANADA
liane.gabora@ubc.ca

**Abstract**

EVOC (for EVOlution of Culture) is a computer model of culture that enables us to investigate how various factors such as barriers to cultural diffusion, the presence and choice of leaders, or changes in the ratio of innovation to imitation affect the diversity and effectiveness of ideas. It consists of neural network based agents that invent ideas for actions, and imitate neighbors' actions. The model is based on a theory of culture according to which what evolves through culture is not memes or artifacts, but the internal models of the world that give rise to them, and they evolve not through a Darwinian process of competitive exclusion but a Lamarckian process involving exchange of innovation protocols. EVOC shows an increase in mean fitness of actions over time, and an increase and then decrease in the diversity of actions. Diversity of actions is positively correlated with population size and density, and with barriers between populations. Slowly eroding borders increase fitness without sacrificing diversity by fostering specialization followed by sharing of fit actions. Introducing a leader that broadcasts its actions throughout the population increases the fitness of actions but reduces diversity of actions. Increasing the number of leaders reduces this effect. Efforts are underway to simulate the conditions under which an agent immigrating from one culture to another contributes new ideas while still 'fitting in'.

## Introduction

What impact do leaders and role models have on the views and behaviors in a given culture? Is a dictatorial or a distributed mode of leadership more effective? What is the effect of complete or semi-permeable barriers to trade and immigration, or barriers that erode or strengthen with time? And what implications do these kinds of cultural patterns have on how best to negotiate, conduct business, or simply behave in a foreign land? These questions and others are addressed using a computer model of culture referred to as EVOC (for EVOlution of Culture). EVOC consists of neural network based agents that invent ideas for actions, and imitate neighbors' actions (Gabora, 2008). EVOC is an elaboration of Meme and Variations, or MAV (Gabora, 1994, 1995), the earliest computer program to model culture as an evolutionary process in its own right. MAV was inspired by the genetic algorithm (GA), a search technique that finds solutions to complex problems by generating a 'population' of candidate solutions through processes akin to mutation and recombination, selecting the best, and repeating until a satisfactory solution is found. Although MAV has inspired the incorporation of cultural phenomena (such as imitation, knowledge-based operators, and mental simulation) into evolutionary search algorithms (e.g. Krasnogor & Gustafson, 2004), the goal behind MAV was not to solve search problems, but to gain insight into how ideas evolve. It used neural network based agents that could (1) invent new ideas by modifying previously learned ones, (2) evaluate ideas, (3) implement ideas as actions, and (4) imitate ideas implemented by neighbors. Agents evolved in a cultural sense, by generating and sharing ideas for actions, but not in a biological sense; they neither died nor had offspring. The approach can thus be contrasted with computer models of the interaction between biological evolution and individual learning (Best, 1999, 2006; Higgs, 2000; Hinton & Nowlan, 1987; Hutchins & Hazelhurst, 1991).

MAV successfully modeled how 'descent with modification' can occur in a cultural context, but it had limitations arising from the outdated methods used to program it. Moreover, although new ideas in MAV were generated making use of acquired knowledge and pattern detection, the name 'Meme and Variations' implied acceptance of the notion that cultural novelty is generated randomly, and that culture evolves through a Darwinian process operating on discrete units of culture, or 'memes'. Problems with memetics and other Darwinian approaches to culture have become increasingly apparent (Boone & Smith, 1998; Fracchia & Lewontin, 1999; Gabora, 2004, 2006, 2008; Jeffreys, 2000). One problem is that natural selection prohibits the passing on of acquired traits (thus you don't inherit your mother's tattoo).[1] In culture, however, 'acquired' change—that is, modification to ideas between the time they are learned and the time they are expressed—is unavoidable. Darwinian approaches must assume that elements of culture are expressed in the same form as that in which they are acquired. Natural selection also assumes that lineages do not intermix. However, because ideas cohabit a distributed memory with a multitude of other ideas, they are constantly combining to give new ideas, and their meanings, associations, and implications are constantly revised.

[1] That isn't to say that inheritance of acquired traits never occurs in biological evolution; it does. However to the extent that this is the case natural selection cannot provide an accurate model of biological evolution. Because inheritance of acquired traits is the exception in biology not the rule, natural selection still provides a roughly accurate model of biological evolution.

It has been proposed what evolves through culture is not discrete memes or artifacts, but the internal models of the world that give rise to them (Gabora, 2004), and they evolve not through a Darwinian process of competitive exclusion but a Lamarckian process involving exchange of innovation protocols (Gabora, 206, 2008). EVOC incorporates this in part by allowing agents to have multiple interacting needs, thereby fostering complex actions that fulfill multiple needs. Elsewhere (Gabora, 2008) the results of experiments using different needs and/or multiple needs are described.

This paper describes other experiments carried out with EVOC that were not possible to carry out with MAV. These experiments investigate how cultural evolution is affected by leadership, and by affordances of the agents' world, such as world shape and size, population density, and barriers that impede information flow, and potentially erode with time.

## Architecture

EVOC consists of an artificial society of agents in a two-dimensional grid-cell world. It is written in Joone, an object oriented programming environment, using an open source neural network library written in Java. This section describes the key components of the agents and the world they inhabit.

### The Agent

Agents consist of (1) a neural network, which encodes ideas for actions and detects trends in what constitutes a fit action, and (2) a body, which implements actions. In MAV there was only one need—to attract a mate. Thus actions were limited to gestures that attract mates. In EVOC agents can also engage in tool-making actions.

**The Neural Network**. The core of an agent is a neural network, as shown in Figure 1. It is composed of six input nodes that represent concepts of body parts (LEFT ARM, RIGHT ARM, LEFT LEG, RIGHT LEG, HEAD, and HIPS), six matching output nodes, and six hidden nodes that represent more abstract concepts (LEFT, RIGHT, ARM, LEG, SYMMETRY and MOVEMENT). Input nodes and output nodes are connected to 'hidden' nodes of which they are instances (e.g. RIGHT ARM is connected to RIGHT.) Activation of any input node increases activation of the MOVEMENT hidden node. Opposite-direction activation of pairs of limb nodes (e.g. leftward motion of one arm and rightward motion of the other) activates the SYMMETRY node.

The neural network learns ideas for actions. An idea is a pattern of activation across the output nodes consisting of six elements that instruct the placement of the six body parts. Training of the neural network is as per (Gabora, 1995). In brief, the neural network starts with small random weights, and patterns that represent ideas for actions are presented to the network. Each time a pattern is presented, the network's actual output is compared to the desired output. An error term is computed, which is used to modify the pattern of connectivity in the network such that its responses become more correct. Since the neural network is an autoassociator, training continues until the output is identical to the input. At this point training stops and the run begins. The value of using a neural network is simply that trends about what makes for a fit action can be detected using the symmetry and movement nodes (see below). The neural network can also be turned off to compare results to those obtained using instead of a neural network a simple data structure that cannot detect trends, and thus invents ideas at random.

**Knowledge-based Operators**. Brains detect regularity and build schemas with which they adapt the mental equivalents of mutation and recombination to tailor actions to the situation at hand. Thus they generate novelty strategically, on the basis of past experience. Knowledge-based operators are a crude attempt to incorporate this into the model. Since a new idea for an action is not learned unless it is fitter than the currently implemented action, newly learned actions provide valuable information about what constitutes an effective idea. This information is used by knowledge-based operators to probabilistically bias invention such that new ideas are generated strategically as opposed to randomly. Thus the idea is to translate knowledge acquired during evaluation of an action into educated guesses about what makes for a fit action.

Two rules of thumb are used. The first rule is: if movement is generally beneficial, the probability increases that new actions involve movement of more body parts. Each body part starts out at a stationary rest position, and with an equal probability of changing to movement in one direction or the other. If the fitter action codes for more movement, increase the probability of movement of each body part. Do the opposite if the fitter action codes for less movement. This rule of thumb is based on the assumption

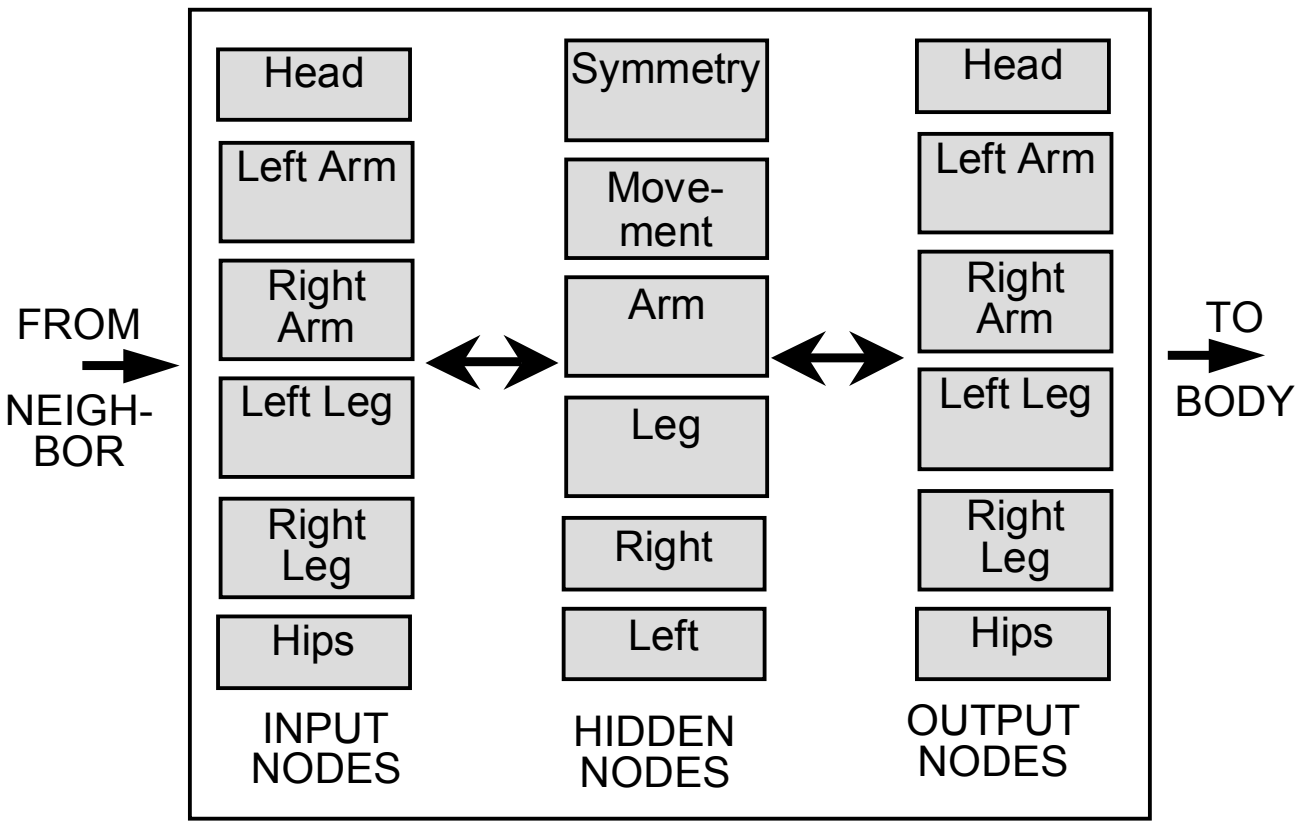

Figure 1. The neural network. See text for details.

that movement in general (regardless of which particular body part is moving) can be beneficial or detrimental. This seems like a useful generalization since movement of any body part uses energy and increases the likelihood of being detected. It is implemented as follows:

$a_{m1}$ = movement node activation for current action
$a_{m2}$ = movement node activation for new action
$p(im)_i$ = probability of increased movement at body part $i$
$p(dm)_i$ = probability of decreased movement at body part $i$

IF ($a_{m2} > a_{m1}$)
THEN $p(im)_i$ = MAX(1.0, $p(im)_i$ + 0.1)
ELSE IF ($a_{m2} < a_{m1}$)
THEN $p(im)_i$ = MIN(0.0, $p(im)_i$ - 0.1)
$p(dm)_i = 1 - p(im)_i$

The second rule of thumb is: if fit actions tend to be symmetrical (e.g. left arm moves to the right and right arm moves to the left), the probability increases that new actions are symmetrical. This generalization is biologically sensible, since many useful actions (e.g. walking) entail movement of limbs in opposite directions, while others (e.g. pushing) entail movement of limbs in the same direction. This rule is implemented in a manner analogous to that of the first rule.

In summary, each action is associated with a measure of its effectiveness, and generalizations about what seems to work and what does not are translated into guidelines that specify the behavior of the algorithm.

**The Body**. If the fitness of an action is evaluated to be higher than that of any action learned thus far, it is copied from the output nodes of the neural network that represent *concepts of* body parts to a six digit array that contains represenions of *actual* body parts, referred to as the *body*. Since it is useful to know how many agents are doing essentially the same thing, when node activations are translated into limb movement they are thresholded such that there are only three possibilities for each limb: stationary, left, or right. Six limbs with three possible positions each gives a total of 729 possible actions. Only the action that is currently implemented by an agent's body can be observed and imitated by other agents.

## The Fitness Functions

Agents evaluate the effectiveness of their actions according to how well they satisfy needs using a pre-defined equation referred to as a *fitness function*. Agents have two possible needs. The fitness of an action with respect to the need to attract mates is referred to as $F_1$, and it is calculated as in (Gabora, 1995). $F_1$ rewards actions that make use of trends detected by the symmetry and movement hidden nodes and used by knowledge-based operators to bias the generation of new ideas. $F_1$ generates actions that are relatively realistic mating displays, and exhibits a cultural analog of *epistasis*. In biological epistasis, the fitness conferred by the allele at one gene depends on which allele is present at another gene. In this cognitive context, epistasis is present when the fitness contributed by movement of one limb depends on what other limbs are doing.

The fitness of an action with respect to the second need, the need to make tools, uses a second fitness function, $F_2$, and is calculated as in (Gabora, 2008).

## Incorporation of Cultural Phenomena

In addition to knowledge-based operators, discussed previously, agents incorporate the following phenomena characteristic of cultural evolution as parameters that can be turned off or on (in some cases to varying degrees):

- **Imitation**. Ideas for how to perform actions spread when agents copy neighbors' actions. This enables them to share effective, or 'fit', actions.
- **Invention**. This code enables agents to generate new actions by modifying their initial action or a previously invented or imitated action using knowledge-based operators (discussed previously).
- **Mental simulation**. Before committing to implementing an idea as an action, agents can use the fitness function to assess how fit the action would be if it *were* implemented.

## The World

MAV allowed only worlds that were toroidal, or 'wrap-around'. Moreover, the world was always maximally densely populated, with one agent per cell. In EVOC the world can be either toroidal or square, and as sparsely or densely populated as desired, with agents placed in any configuration. EVOC also allows the creation of complete or semi-permeable permanent or eroding borders that decrease the probability of imitation along a frontier.

## A Typical Run

Each iteration, every agent has the opportunity to (1) acquire an idea for a new action, either by *imitation*, copying a neighbor, or by *invention*, creating one anew, (2) update the knowledge-based operators, and (3) implement a new action. To invent a new idea, the current action is copied to the input layer of the neural network, and this previous action is used as a basis from which to generate a new one. For each node the agent makes a probabilistic decision as to whether change will take place. If it does, the direction of change is stochastically biased by the knowledge-based operators using the activations of the SYMMETRY and MOVEMENT nodes. Mental simulation is used to determine whether the new idea has a higher fitness than the current action. If so, the agent learns and implements the action specified by the new idea.

To acquire an idea through imitation, an agent randomly chooses one of its neighbors, and evaluates the fitness of the action the neighbor is implementing using mental simulation. If its own action is fitter than that of the neighbor, it chooses another neighbor, until it has either observed all of its immediate neighbors, or found one with a fitter action. If no fitter action is found, the agent does nothing. Otherwise, the neighbor's action is copied to the input layer, learned, and implemented.

Fitness of actions starts out low because initially all agents are immobile. Soon some agent invents an action that has a higher fitness than doing nothing, and this action gets imitated, so fitness increases. Fitness increases further as other ideas get invented, assessed, implemented as actions, and spread through imitation. The diversity of actions initially increases due to the proliferation of new ideas, and then decreases as agents hone in on the fittest actions.

### The Graphical User Interface

The graphical user interface (GUI) makes use of the open-source charting project, JFreeChart, enabling variables to be user defined at run time, and results to become visible as the computer program runs. Figure 2 shows the topmost output panel using the mating fitness function ($F_1$). At the upper left one specifies the *Invention to Imitation Ratio*. This refers to the probability that a given agent, on a given iteration, invents a new idea for an action, versus the probability that it imitates a neighbor's action. Below that is *Rate of Conceptual Change*, where one specifies the degree to which a newly invented idea differs from the one it was based on. Below that is *Number of Agents*, which allows the user to specify the size of the artificial society. Below that is where one specifies *Number of Iterations*, i.e. the duration of a run. The agents that make up the artificial society can be accessed individually by clicking the appropriate cell in the grid on the upper right. This enables one to see such details as the action currently implemented by a particular agent, or the fitness of that action. The graphs at the bottom plot the mean idea fitness and diversity of ideas. Tabs shown at the top give access to other output panels of the GUI.

## Summary of Previous Results

EVOC closely replicates the results of experiments conducted with MAV (Gabora, 1995). The graph on the bottom left of Figure 2 shows the increase in fitness of actions. The graph on the bottom right of Figure 2 shows the increase and then decrease in the diversity of actions. Other MAV results that are replicated with EVOC include:

- Fitness increases most quickly with an invention to imitation ratio of approximately 2:1.
- For the agent with the fittest actions, however, the *less* it imitates, the better it does.
- Increasing the invention-to-imitation ratio increases the diversity of actions. If increased much beyond 2:1, it takes more than twice as many iterations for all agents to settle on optimal actions.
- As in biology, epistatically linked elements take longer to optimize. (As explained earlier, in the present context epistasis refers to the situation where the effect on fitness of what one limb is doing depends on what another is doing.)
- The program exhibits *drift*—the term biologists use to refer to changes in the relative frequencies of alleles (forms of a gene) as a statistical byproduct of randomly sampling from a finite population (Wright, 1969). With respect to culture, the term pertains not to alleles but to possible forms of a component of an idea (e.g. if the idea is to implement the gesture 'wave', one can do this with one's left hand or one's right).

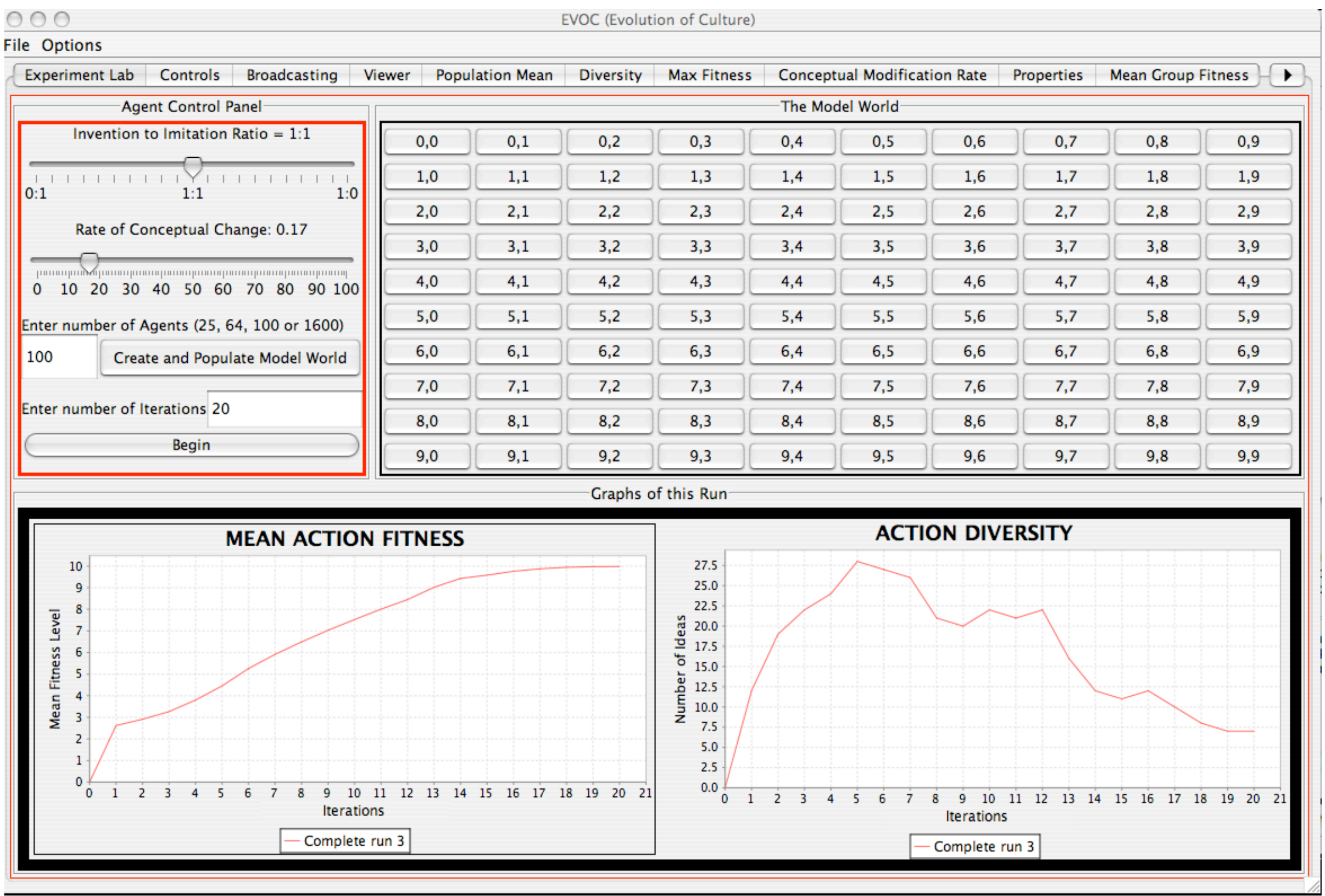

Figure 2. Output panel of GUI using $F_1$. See text for details.

These results show that concepts from biology are useful in the analysis of cultural change, but that culture also exhibits phenomena that have no biological equivalent.

Previous work on EVOC focused on the effects of changing the need, and integrating multiple needs (Gabora, 2008). It was found that changing the need (modeled as a change in the fitness function) or giving agents multiple needs to fulfill, does not change the overall pattern of results. Mean fitness of actions still increases gradually, and diversity of actions rises and then falls, exhibiting the typical inverted U-shaped curve, the magnitude of which is a function of population size. However, increasing the number of needs consistently results in a higher diversity of actions, and tends to decrease mean fitness with respect to any given need.

## Experiments

We now outline the results of current experiments with EVOC. Unless stated otherwise, the world is toriodal and consists of 100 cells, with maximum density (one agent per cell), no broadcasting, no barriers to idea flow, a 1:1 invention to imitation ratio, and a 0.17% probability of change to any body part during invention (since, with six body parts, on average each newly invented action differs from the one it was based on with respect to one body part).

### Complete and Semi-permeable Barriers

Throughout history, the flow of ideas has been impeded geographical barriers and political/cultural borders. It is possible to simulate this in EVOC by reducing the probability of imitation between agents on opposite sides of a barrier. Barriers were found to increase latency to converge on fit actions, and to increase diversity, by effectively dividing the population. Interesting results are achieved when barriers erode over time such that the probability of imitation by agents on opposite sides is initially zero but increases over the duration of a run, simulating globalization. Figure 3 shows the diversity of actions implemented after 4 iterations with an eroding barrier. Eroding barriers foster specialization—honing in on unique solutions—on different sides of the border, followed by sharing of the best to reach a diverse final set.

| 0,0 | 0,1 | 0,2 | 0,3 | 0,4 | 0,5 | 0,6 | 0,7 |
|---|---|---|---|---|---|---|---|
| 1,0 | 1,1 | 1,2 | 1,3 | 1,4 | 1,5 | 1,6 | 1,7 |
| 2,0 | 2,1 | 2,2 | 2,3 | 2,4 | 2,5 | 2,6 | 2,7 |
| 3,0 | 3,1 | 3,2 | 3,3 | 3,4 | 3,5 | 3,6 | 3,7 |
| 4,0 | 4,1 | 4,2 | 4,3 | 4,4 | 4,5 | 4,6 | 4,7 |
| 5,0 | 5,1 | 5,2 | 5,3 | 5,4 | 5,5 | 5,6 | 5,7 |
| 6,0 | 6,1 | 6,2 | 6,3 | 6,4 | 6,5 | 6,6 | 6,7 |
| 7,0 | 7,1 | 7,2 | 7,3 | 7,4 | 7,5 | 7,6 | 7,7 |

Figure 3. Diversity of actions after four iterations with 8x8 grid and eroding barrier between 3rd and 4th columns. Different actions represented by different colored cells (which will appear in print as different shades of grey). Invention to imitation ratio of agents to right of border twice is that of agents to the left.

Diversity similarly affected by whether the shape of the world is square – which simulates a situation where the flow of ideas is bounded by natural or political boundaries – versus toroidal -- which simulates the situation where such barriers are overcome through globalization. A global (toroidal) world accelerates fitness and increases diversity in the short term but decreases it in the long term. This makes sense; agents at the edges of a square world have fewer neighbors, and thus more opportunity to retain deviant actions.

### Effect of Population Density

EVOC allows not just the shape of the world to be changed, but how densely populated it is. Figure 4 illustrates the diversity of actions over a run with different population densities.

The lower the population density, the more the typical inverted-U shaped action diversity curve is disrupted; both the initial peak and subsequent decline are less pronounced. Further analysis reveals that decreasing population density fosters the existence of small isolated clusters that are unable to learn from one another and share effective actions, impairing the ability of the society to converge on only the fittest actions.

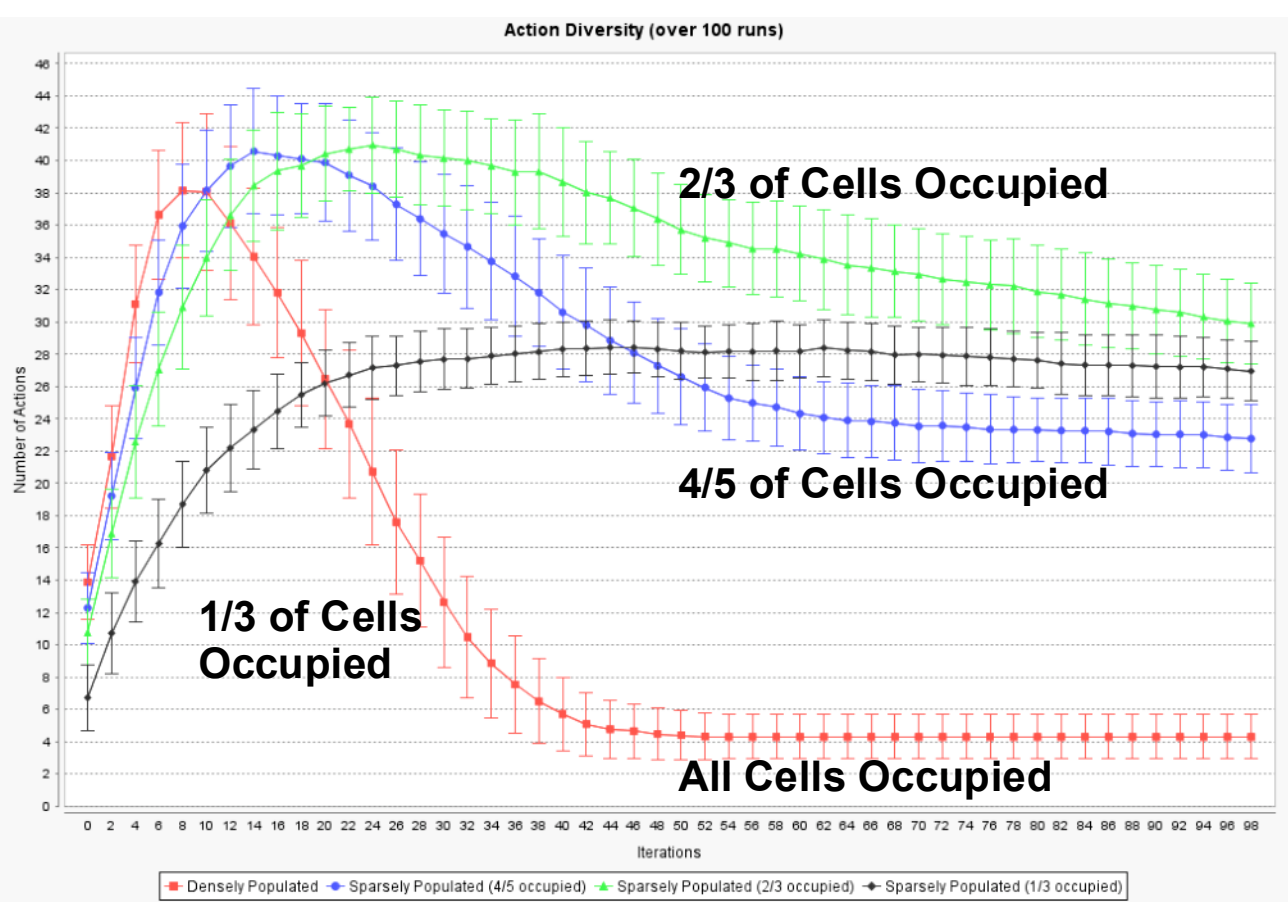

Figure 4. Effect of varying population density on diversity of actions.

### Broadcasting

Broadcasting allows the action of a leader, or broadcaster, to be visible to not just immediate neighbors, but all agents, thereby simulating the effects of media such as public performances, television, radio, or internet, on patterns of cultural change. When broadcasting is turned on, an agent is no longer limited to its immediate neighbors as potential role models. Each agent adds the broadcaster as a possible source of actions it can imitate. A particular agent can be chosen as the broadcaster before the run, or the broadcaster can be chosen at random, or the user can specify that the agent with the fittest action is the broadcaster. Broadcasting can be intermittent, or continued throughout the duration of a run. Broadcasting does not have a significant effect on the fitness of actions, but as shown in Figure 5, it accelerates convergence on optimal actions, and consistently reduces diversity.

By varying the number of broadcasters EVOC allows simulation of the effect on fitness and diversity of ideas of a dictorial style of leadership (one broadcaster) versus a distributed style of leadership (multiple leaders). In figure 6 we see how adding the presence of a broadcaster (comparing column 1 without broadcaster to column 2 with broadcaster) decreases the diversity of actions. This i seen clearly looking to the lowest row: 20 iterations. Whereas without a broadcaster there are eight different actions, and 41% of agents are executing the most popular action, with a broadcaster there are five different actions, and 84% of agents are executing the most popular action. However, as shown in columns 3 to 6, these trends become increasingly reversed the greater the number of broadcasters. With five broadcasters, the society converges on nine different actions, and the percentage of agents executing the most popular action is down to 31%. These data potentially speak to the changing effect of media on society. With only one television or radio station, the effect of media may have been to make opinions and behaviors more homogeneous. However with the proliferation of different radio and television stations, as well as web-based media, available from not just local sources but around the world, the effect might well be the reverse: an explosion of different views and behaviors.

## Discussion

This paper has given an overview of factors impacting the spread of ideas and behaviors that can be investigated with a computer model of cultural evolution, focusing on new results investigating the effects of broadcasting (leadership), population density, and the shape and penetrability (e.g. presence of boundaries) of the terrain. Results suggest that properties of the world can have as great an impact on the evolution of culture as properties of the agents themselves. The results also show that the benefits of leadership with respect to enhanced fitness of ideas may be tempered by decreased diversity of ideas. This echoes previous simulation findings that leadership can have adverse effects when agents can communicate (Gigliotta, Miglino, & Parisi, 2007).

A primary aim of future work will be to examine the distinctively human phenomenon of cultural open-endedness. Although presently agents' actions become more complex and adapted over time, and change is cumulative in that new actions build on existing ones, once agents settle on some subset of optimal actions, the program comes to a standstill. Future versions will use a fitness function that evaluates actions differently depending on the relative strengths of the different needs. The strength of a need will be a function of both how many iterations have passed since execution of an action that satisfied that need, and the degree to which that action satisfied that need. It is expected that the program will not come to a standstill because once an agent has filled one need it will change the kind of action it implements to satisfy another. Moreover to avoid that agents still zero in on predictable subsets of actions that fulfill these needs, future versions of EVOC will incorporate the following:

- ***Context-sensitive concepts***. We plan to move to a more subsymbolic level, incorporating how constellations of activated microfeatures are influenced by context (Aerts & Gabora, 2005a,b; Gabora, Rosch, & Aerts, 2008). This will allow for a richer repertoire of actions.
- ***Chained Actions***. Agents will be allowed to chain actions into arbitrarily long action sequences.
- ***Building Blocks***. Agents will implement actions that

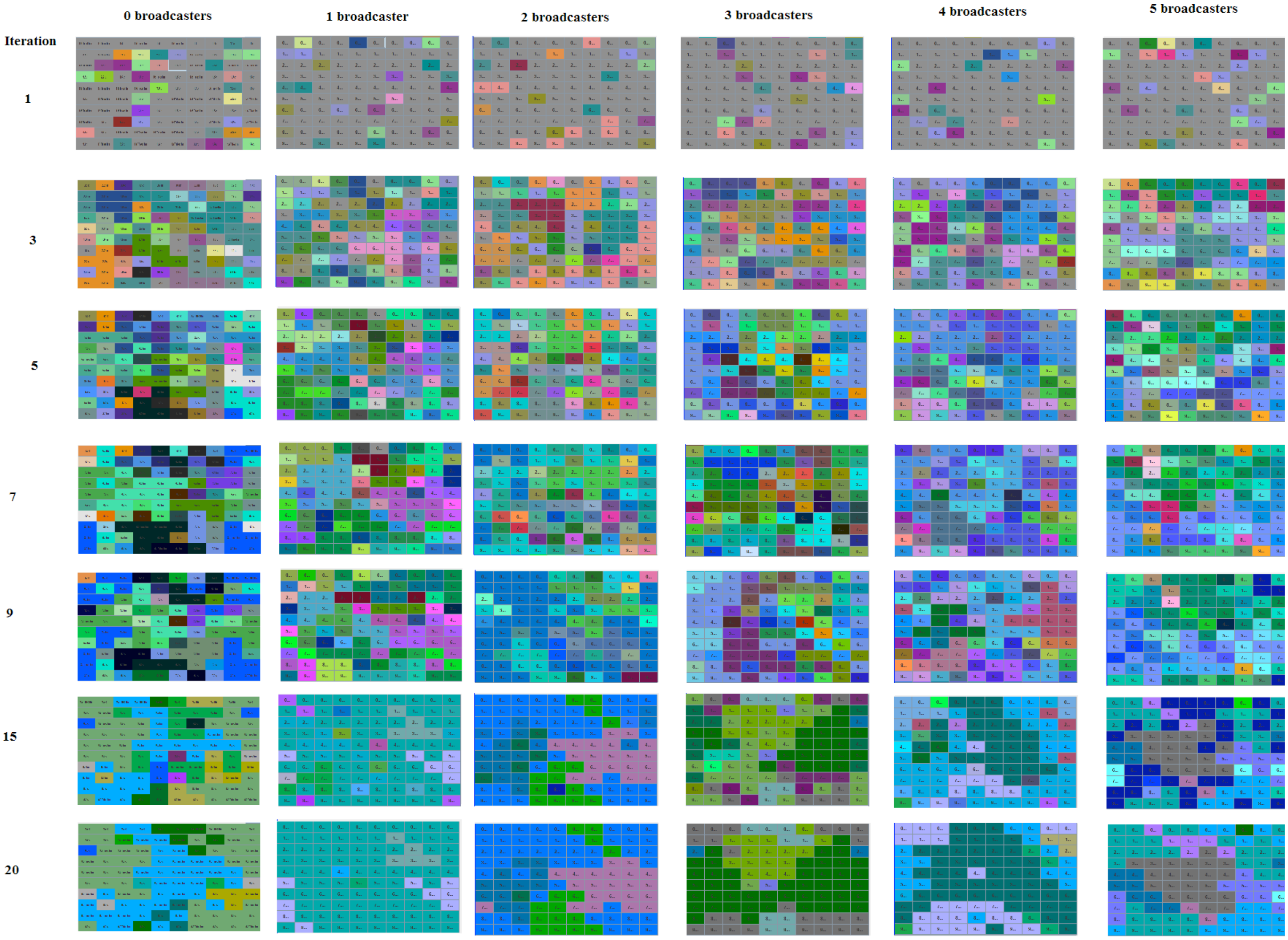

Figure 6. Diversity of actions over a run with 0, 1, 2, 3, 4, and 5 broadcasters. Different actions are represented by differently colored cells (which will appear in the printed version as different shades of grey). In all cases there is an increase followed by a decrease in diversity over time (moving down any column from the first iteration, at the top, to the 20th iteration, at the bottom). However, the decrease is less pronounced the more broadcasters there are. These runs used a toroidal, maximally dense world, with a 10x10 grid. In this run, broadcasters were chosen at random every iteration, and when their were multiple broadcasters, agents selected the broadcaster whose action was most similar to their own.

cumulatively modify their world using building blocks to create structures that satisfy needs, and add to (or destroy) structures made by others.

With these modifications it is expected that there will no longer be an *a priori* limit to the number or complexity of actions. The role of each of these modifications in bringing about genuine cultural evolution will be assessed. The effort will be judged successful if cultural change is not just cumulative, but cumulative in a way that responds to needs and situations, and open-ended, such that one innovation creates niches for the invention of others (as cars paved the way for seat belts and gas stations).

Further experiments with eroding barriers has potential implications for the impact of free trade on global diversity of ideas, and for investigating the complex relationship between creativity and culture (Kaufman & Sternberg, 2006). Future efforts will also focus on a more in-depth analysis of the conditions under which immigrant contributes to the fitness and diversity of ideas versus the conditions under which the immigrant's actions are so different that they merely stand out and do not contribute in a productive way. A wider range of needs will be made available in order to determine the relationship between degree of similarity between needs of the native and immigrant populations and latency of the immigrant population to 'fit in'. Other questions will also be investigated, such as 'How does the probability of 'fitting in' change as a function of the number of immigrants?

These issues are timely, and have potential implications for how individuals should go about negotiating, conducting business, and simply behaving in a foreign land. Even when such simulations do not provide specific directives, they help us to think in more precise terms about the issues.

## Acknowledgments

Thanks to Martin Denton and Jillian Dicker for their work on EVOC. This project is funded by the Social Sciences and Humanities Research Council of Canada (SSHRC).